\documentclass[english]{sbrt}
\usepackage[utf8]{inputenc}

\usepackage{cite}
\usepackage{placeins}
\usepackage{amsmath,amssymb,amsfonts}
\usepackage{algorithmic}
\usepackage{graphicx}
\usepackage{textcomp}
\usepackage{xcolor}
\usepackage{listings}
\usepackage[none]{hyphenat}
\usepackage{url}
\usepackage{fancyvrb}
\usepackage{fvextra}
\usepackage{comment}
\usepackage{subcaption} 
\usepackage{makecell}
\usepackage[english]{babel}
\usepackage{minted}
\setminted[bash]{  
    fontsize=\footnotesize,
    breaklines=true,
    breakafter=/,
    breaksymbolleft={\tiny\ensuremath{\hookrightarrow}},
    frame=leftline,
    framesep=2mm,
    bgcolor=lightgray!5,
    autogobble=true,
    tabsize=2
}
\usepackage{tabularx}
\usepackage{pifont} 

\usepackage[acronym,toc,shortcuts]{glossaries}
\newacronym{ufrn}{UFRN}{Universidade Federal do Rio Grande do Norte}
\newacronym{ccet}{CCET}{Centro de Ciências Exatas e da Terra}
\newacronym{dimap}{DIMAp}{Departamento de Informática e Matemática Aplicada}
\newacronym{ppgsc}{PPgSC}{Programa de Pós-graduação em Sistemas e Computação}

\newacronym{3gpp}{3GPP}{3rd Generation Partnership Project}
\newacronym{etsi}{ETSI}{European Telecommunications Standards Institute}
\newacronym{itu-t}{ITU-T}{Telecommunication Standardization Sector}
\newacronym{ieee}{IEEE}{Institute of Electrical and Electronics Engineers}

\newacronym{cn}{CN}{Core Network}
\newacronym{embb}{eMBB}{Enhanced Mobile Broadband}
\newacronym{mmtc}{mMTC}{Massive Machine Type Communication}
\newacronym{urllc}{URLLC}{Ultra-reliable and Low Latency Communications}
\newacronym{v2x}{V2X}{Vehicle-to-Everything}
\newacronym{hmtc}{HMTC}{High-Performance Machine-Type Communications}
\newacronym{hdllc}{HDLLC}{High Data rate and Low Latency Communications}

\newacronym{iot}{IoT}{Internet of Things}
\newacronym{kpi}{KPI}{Key Performance Indicator}
\newacronym{mno}{MNO}{Mobile Network Operator}
\newacronym{naas}{NaaS}{Network as a Service}
\newacronym{nsi}{NSI}{Network Slice Instance}
\newacronym{ran}{RAN}{Radio Access Network}
\newacronym{sdn}{SDN}{Software-Defined Networking}
\newacronym{nfv}{NFV}{Network Functions Virtualisation}
\newacronym{tn}{TN}{Transport Network}
\newacronym{ai}{AI}{Artificial Intelligence}
\newacronym{ml}{ML}{Machine Learning}
\newacronym{cots}{COTS}{Commercial Off-The-Shelf}
\newacronym{ru}{RU}{Radio Unit}
\newacronym{du}{DU}{Distributed Unit}
\newacronym{cu}{CU}{Centralized Unit}
\newacronym{vnf}{VNF}{Virtual Network Functions}
\newacronym{nsst}{NSST}{Network Slice Subnet Template}
\newacronym{oms}{OMS}{Operations and Maintenance System}
\newacronym{mano}{MANO}{Management and Orchestration}
\newacronym{nws}{NWS}{Network Slice}
\newacronym{ns}{NS}{Network Slicing}
\newacronym{ue}{UE}{User Equipment}
\newacronym{pdu}{PDU}{Protocol Data Unit}
\newacronym{plmn}{PLMN}{Public Land Mobile Network}
\newacronym{snpn}{SNPN}{Standalone NPN}
\newacronym{npn}{NPN}{Non-Public Networks}
\newacronym{mvno}{MVNO}{Mobile Virtual Network Operator}

\newacronym{amf}{AMF}{Access and Mobility Management Function}
\newacronym{n3iwf}{N3IWF}{Non-3GPP Interworking Function}
\newacronym{tngf}{TNGF}{Trusted Non-3GPP Gateway Function}
\newacronym{twif}{TWIF}{Trusted WLAN Inter-Working Function}

\newacronym{hetnet}{HetNet}{Heterogeneous Networks}
\newacronym{mptcp}{MPTCP}{MultiPath TCP}
\newacronym{ioms}{iOMS}{Intelligente Orchestration and Management System}
\newacronym{oai}{OAI}{OpenAirInterface}

\newacronym{rat}{RAT}{Radio Access Technology}
\newacronym{rats}{RATs}{Radio Access Technologies}
\newacronym{enb}{eNB}{evolved Node B}
\newacronym{gnb}{gNB}{Next Generation Node B}
\newacronym{lte}{LTE}{Long Term Evolution}
\newacronym{nr}{NR}{New Radio}
\newacronym{ngran}{NG-RAN}{Next Generation Radio Access Network}
\newacronym{wifi}{Wi-Fi}{Wireless Fidelity}
\newacronym{lifi}{Li-Fi}{Light Fidelity}
\newacronym{wlan}{WLAN}{Wireless Local Area Network}
\newacronym{lan}{LAN}{Local Area Network}
\newacronym{ap}{AP}{Access Point}
\newacronym{erb}{ERB}{Estação Rádio Base}
\newacronym{fwa}{FWA}{Fixed Wireless Access}
\newacronym{ftth}{FTTH}{Fiber to the Home}

\newacronym{ar}{AR}{Augmented Reality}
\newacronym{vr}{VR}{Virtual Reality}
\newacronym{mr}{MR}{Mixed Reality}
\newacronym{xr}{XR}{eXtended Reality}
\newacronym{mxr}{MXR}{Multisensory Extended Reality}
\newacronym{b2b}{B2B}{Business to Business}
\newacronym{b2c}{B2C}{Business to Consumer}
\newacronym{atf}{ATF}{Airtime Fairness}
\newacronym{nas}{NAS}{Non-Access Stratum}

\usepackage[normalem]{ulem}
\newif\ifrev
\revtrue 
\ifrev

\else

\newcommand{\rmtexto}[1]{}
\fi

\usepackage[a4paper, left=0.6in, right=0.6in, bottom=0.6in, top=0.7in]{geometry}

\begin{document}
\bstctlcite{IEEEexample:BSTcontrol}
\title{Integrating Wi-Fi into 3GPP 5G Network Slicing: An Experimental Prototype Study}

\author{Nelson Ion de Oliveira, Marília Costa Muniz, William M. C. do Nascimento, João Pedro Brasil, Victor Farias Monteiro, Sérgio Barros, Maykon Silva, Augusto Venâncio Neto e Vicente A. de Sousa Jr.
\thanks{Sérgio Barros and Maykon Silva (E-mails: \{sbarros,mpsilva\}@cpqd.com.br) are from Telecommunication Research and Development Center (CPQD). This work is partially supported by CPQD and Ministry of Science, Technology and Innovation (MCTI), with financial resources from FUNTTEL and administered by FINEP, specifically within the scope of the projects AERF - Strategic Actions for Future Networks, Contract 01.22.0471.01, Reference 1508/22. Victor Monteiro is from the Federal University of Ceará (victor@gtel.ufc.br). João Brasil is a researcher from Instituto Atlântico (joao\_brasil@atlantico.com.br).  They are supported by FINEP and MCTI, through the project \textit{AITD LITEC: Avanços em Inteligência Artificial para Transformação Digital no Laboratório de Inovação Tecnológica e Experimentos Científicos} (Ref. FINEP 1039/24, Grant nº 01.24.0581.00). Victor Monteiro is also supported by CNPq under the  Grant DT-308267/2022-2.  The remaining authors are with the Leading Advanced Technologies Center of Excellence (LANCE) at the Federal University of Rio Grande do Norte. E-mails: nelson@imd.ufrn.br, marilia.muniz.124@ufrn.edu.br, william.nascimento.086@ufrn.edu.br, augusto@dimap.ufrn.br, vicente.sousa@ufrn.br. This work is also supported by Coordenação de Aperfeiçoamento de Pessoal de Nível Superior - Brazil (CAPES) - financing code 001 and by \textit{Instituto Nacional de Ciência, Tecnologia e Inovação} (INCT) de Redes Inteligentes de Comunicações e Internet das Coisas (ICoNIoT) (CNPq Grant No. 405940/2022-0 e CAPES Grant No. 88887.954253/2024-00). We attest that generative AI (Copilot, Gemini, and NotebookLM) was used for linguistic editing, text refinement, and scripts coding.}}

\maketitle

\markboth{XLIV SIMPÓSIO BRASILEIRO DE TELECOMUNICA\c{C}\~{O}ES E PROCESSAMENTO DE SINAIS - SBrT 2026, 29 DE SETEMBRO A 2 DE OUTUBRO DE 2026, SALVADOR, BAHIA}{}

\begin{abstract}
Network Slicing (NS) is a fundamental pillar of 5G and beyond networks, enabling the provisioning of isolated, logical networks tailored to specific Quality of Service (QoS) requirements. While 3GPP standards comprehensively define slicing architectures over cellular access networks, the seamless integration of Non-3GPP technologies such as Wi-Fi into a unified slice instance remains an active area of investigation, particularly regarding empirical validation. This paper presents an end-to-end prototyping study that integrates 5G Standalone (SA) and Wi-Fi networks by adapting the Trusted Non-3GPP Gateway Function~(TNGF) to extend NS to WLAN networks, enabling the unified management of Wi-Fi transmission resources. We implement a functional testbed leveraging an open-source 5G Core and an explicit Non-3GPP access to validate multi-Radio Access Technology (multi-RAT) slice operation. Our empirical results showcase the dynamic viability of multi-RAT slicing under varying bandwidth allocations and traffic steering policies, providing a concrete proof of concept for unified 3GPP and Non-3GPP service delivery.
\end{abstract}



\section{Introduction}

The evolution toward fifth-generation (5G) and sixth-generation (6G) cellular systems marks a paradigm shift from rigid, single-service architectures to highly flexible, service-oriented networks. Network Slicing (NS) \cite{3gpp_ts28530_v18} is the primary technology enabling this transition, allowing network operators to partition a physical infrastructure into multiple virtual networks optimized for diverse use cases such as ultra-reliable low-latency communications (URLLC), enhanced mobile broadband (eMBB), and massive machine-type communications (mMTC) \cite{3gpp23502_v20_1}. 

To fulfill the vision of truly ubiquitous, cost-effective, and high-capacity connectivity, Network Slice Instances (NSI) must extend beyond traditional 3GPP Radio Access Technologies (RATs) \cite{3gpp_ts_23_501_v19_9_2025}: 5G New Radio (NR) and 4G Long Term Evolution (LTE). Non-3GPP access networks, primarily Wi-Fi, carry an overwhelming majority of local and indoor data traffic worldwide. We depart from the hypothesis that seamless integration of Wi-Fi (Non-3GPP) into a 4G/5G (3GPP) NSI enables mobile operators to effectively deliver diverse service verticals, such as Extended Reality (XR) and the Internet of Things (IoT), in which the majority of devices are Wi-Fi-based. Aside from this, operators can also leverage existing Wi-Fi dense footprints to offload traffic, boost indoor coverage, and guarantee end-to-end Service Level Agreements (SLAs) without incurring prohibitive capital expenditures.

While the 3GPP architecture defines abstract frameworks for Non-3GPP access through the Trusted Network Gateway Function (TNGF) \cite{3gpp23502_v20_1}, most contemporary research relies heavily on mathematical modeling or software simulations (e.g., ns-3, Mininet). While useful, these simulation-based insights often overlook real-world signaling details, encapsulation overheads, and processing latencies inherent to a real virtualized infrastructure. To bridge this gap, this paper puts forward an end-to-end, multi-RAT network slicing open-source testbed with a real Non-3GPP technology protocol stack. The key contributions of this prototyping study are:

\begin{itemize}
    \item A concrete implementation of trusted Non-3GPP access using a TNGF module to establish secure IPsec tunnels from UEs across a Wi-Fi access network directly into the core user plane.
    \item Extension of 5G benefits to legacy devices, allowing devices without native 5G support to benefit from advanced network features, such as logical isolation, service differentiation, and traffic management, while connected via WLAN.
    \item Comprehensive empirical Proof of Concept (PoC) that showcases real-time multi-RAT NSI operation, along with observation of throughput impacts through dynamic bandwidth adaptation under distinct traffic conditions.
\end{itemize}

The remainder of this paper is structured as follows. Section \ref{sec_trabalhos_relacionados} reviews related work. Section \ref{sec_prototype} details the multi-RAT network slice solution and experimental setup. Section~\ref{sec_setup} presents experimental results and performance metrics, followed by results and discussion in Section~\ref{sec_results}. Section~\ref{sec_conclusoes} concludes the paper.

\section{Related Work}
\label{sec_trabalhos_relacionados}
We conducted a literature review on multi-RAT network slicing and Wi-Fi integration, covering the Scopus and Web of Science databases. The review protocol was based on the following search string: \textit{(802.11 OR WiFi OR Wi-Fi OR wireless) AND (network slic*) AND (wlan OR lwa OR lwip)}. This search returned 136 records, which were analyzed to identify the main strategies investigated in the literature.

The selection process was carried out in two stages. In the first stage, titles, keywords, abstracts, and conclusions were screened, resulting in the selection of 35 papers for full-text reading. In the second stage, a thorough analysis of these papers was performed, focusing on the proposed Wi-Fi slicing techniques, their integration with 5G networks, the technologies employed, and the validation methodologies used. This review, complemented by the snowballing technique applied to both scientific publications and standardization documents, served as the foundation for understanding the current state of the art in Wi-Fi network slicing and its integration with 3GPP networks. 

A synthesis of the main findings from the analyzed works is presented in Table \ref{tab:sintese-trabalhos-relacionados}. Of the 35 papers selected for full-text reading, only 23 specifically addressed network slicing in Wi-Fi networks.

\begin{table}[ht]
\centering
\scriptsize
\caption{Summary Table of Related Works.}
\label{tab:sintese-trabalhos-relacionados}
\begin{tabularx}{\columnwidth}{|p{1.2cm}|X|X|X|}
\hline
\textbf{Ref.} & \textbf{Focus} & \textbf{Results} & \textbf{Limitations} \\ \hline

\cite{9297339} & Slicing for IoT in Wi-Fi & Throughput and QoS improvement & No integration with 5G \\ \hline
\cite{9110407} & Dynamic airtime allocation & QoS guarantees under congestion & Simulation-only, without 5G \\ \hline
\cite{9124869} & WLAN slicing with SDN & Real-time QoS & No integration with 5G \\ \hline
\cite{9155532} & Slicing in SD-RAN & Wi-Fi and LTE integration & No specific focus on 5G \\ \hline
\cite{9129831} & QoS-aware slicing & QoS under limited resources & Wi-Fi only \\ \hline
\cite{9463988} & Slicing with multiple SSIDs & Latency and PER improvement & No integration with 5G \\ \hline
\cite{9373012} & WLAN + 5G for healthcare & Improved scheduling & Limited scalability \\ \hline
\cite{Oliveira2021} & QoE fairness & QoE improvement & Wi-Fi only \\ \hline
\cite{9771816} & Dynamic association & Improved slicing for IoT & Without 5G \\ \hline
\cite{9722800} & 5G/Wi-Fi/LiFi integration & Real-time slicing & Limited scalability \\ \hline
\cite{9380999} & Dynamic temporal slicing & Efficient allocation & Partial integration with 5G \\ \hline
\cite{10176206} & Static slicing & Improved PER and latency & Lack of dynamism \\ \hline
\cite{Smida20232175} & Vehicular networks & Dynamic reconfiguration & Without 5G \\ \hline
\cite{Limani2023} & Disaster response & Temporal optimization & Wi-Fi only \\ \hline
\cite{An2023} & WLAN scheduling & Improved resource utilization & Without 5G \\ \hline
\cite{Ahmed2023212} & Wi-Fi HaLow & Improved throughput and energy efficiency & Without 5G \\ \hline
\cite{10257198} & IoT with DRL & QoS improvement & Limited 5G support \\ \hline
\cite{NavarroOrtiz2024} & Industry 4.0 & Up to 1.4 Gbps & 5GC not identified \\ \hline
\cite{10444402} & Band-based slicing & Improved jitter & No focus on 5G \\ \hline
\cite{10539146} & IoT QoS & Improved differentiation & Without 5G \\ \hline
\cite{10638103} & Wi-Fi 6 vs 5G & Higher reliability & Limited detailing \\ \hline
\cite{10620782} & 5G + Wi-Fi & Scalable integration & Limitations poorly explored \\ \hline
\cite{10454645} & Industrial applications & Improved QoS & Without 5G \\ \hline

\end{tabularx}
\normalsize
\end{table}

Network slicing architectures have been investigated in the literature for years. The feasibility of multi-connectivity and traffic splitting across heterogeneous networks has its roots in early LTE-WLAN Aggregation (LWA) and Multi-RAT Dual Connectivity (MR-DC) standards. However, these techniques operate primarily at the RAN layer and do not intrinsically support the end-to-end virtualization and logical isolation mandated by 5G network slicing.

The evaluated works address several techniques applied to Wi-Fi and, in some cases, to 5G, with distinct focuses such as QoS improvements, spectral efficiency, and resource management. However, a recurring limitation was the insufficient integration or lack of collaboration between Wi-Fi and 5G networks. 

Despite these insights, empirical verification on real radio access protocol stacks remains sparse. Existing testbeds often deploy either 5G-only slicing or unsliced multi-RAT setups. This study addresses this exact vacancy by demonstrating the feasibility of establishing an NS instance across WLAN and 5G, enabling a unified evaluation of throughput behavior under real protocol stack deployment constraints.

\section{Prototyping Architecture and Testbed Deployment}
\label{sec_prototype}

This section details the physical and logical architecture of the multi-RAT network slicing testbed. The network design ensures that a single Network Slice Selection Assistance Information (S-NSSAI) identifier spans all 3GPP and Non-3GPP access technologies seamlessly~\cite{3gpp_ts_23_501_v19_9_2025}.

\subsection{Architectural Framework}

The conceptual architecture follows the 3GPP service-based architecture (SBA) for the control plane and an isolated user plane framework. The unified NSI is uniquely identified by an S-NSSAI composed of a Slice/Service Type (SST) and a Slice Differentiator (SD). For the trusted Non-3GPP access control plane, the \gls{tngf} function is employed within the 5GC. The \gls{tngf} is devoted to enabling devices connected to Non-3GPP access networks to communicate with the 5GC through the N2 and N3 interfaces, ensuring the security and integrity of communications \cite{3gpp_ts_23_501_v19_9_2025}.

The architecture of the experimental setup is centered on the free5GC core network implementation. free5GC is an open-source project maintained by the Linux Foundation, with the primary goal of developing a 5G Core (5GC) fully compliant with 3GPP specifications~\cite{free5gc}. Although other open-source platforms also offer a certain level of multi-RAT support, their documentation often lacks sufficient detail to clearly distinguish the specific approaches used in implementing 3GPP-compliant network functions. Each platform adopts distinct implementation strategies and feature sets. Table~\ref{tab:mapa-implementacoes-nfv-por-5gcs} presents a comparative mapping of the network functions implemented by the investigated open-source 5GC platforms.

\begin{table}[htb]
\centering
\scriptsize
\caption{Comparison of network functions available in open-source 5GC solutions.}
\label{tab:mapa-implementacoes-nfv-por-5gcs}
\begin{tabularx}{\columnwidth}{|c|
>{\centering\arraybackslash}X|
>{\centering\arraybackslash}X|
>{\centering\arraybackslash}X|
>{\centering\arraybackslash}X|
>{\centering\arraybackslash}X|}
\hline
 & \textbf{O5GS} & \textbf{f5GC} & \textbf{OAI} & \textbf{Mag} & \textbf{SDC} \\ \hline
AMF   & \checkmark & \checkmark & \checkmark & \checkmark & \checkmark \\ \hline
AUSF  & \checkmark & \checkmark & \checkmark & \checkmark & \checkmark \\ \hline
BSF   & \checkmark & -          & -          & -          & -          \\ \hline
CHF   & -          & \checkmark & -          & -          & -          \\ \hline
LMF   & -          & -          & \checkmark & -          & -          \\ \hline
N3IWF & -          & \checkmark & -          & -          & -          \\ \hline
N3UE  & -          & \checkmark & -          & -          & -          \\ \hline
NEF   & -          & -          & \checkmark & -          & -          \\ \hline
NRF   & \checkmark & \checkmark & \checkmark & -          & \checkmark \\ \hline
NSSF  & \checkmark & \checkmark & \checkmark & -          & \checkmark \\ \hline
NWDAF & -          & -          & \checkmark & -          & -          \\ \hline
PCF   & \checkmark & \checkmark & \checkmark & \checkmark & \checkmark \\ \hline
SCP   & \checkmark & -          & -          & -          & -          \\ \hline
SEPP  & \checkmark & -          & -          & -          & -          \\ \hline
SMF   & \checkmark & \checkmark & \checkmark & \checkmark & \checkmark \\ \hline
TNGF  & -          & \checkmark & -          & -          & -          \\ \hline
TNGFUE& -          & \checkmark & -          & -          & -          \\ \hline
UDM   & \checkmark & \checkmark & \checkmark & \checkmark & \checkmark \\ \hline
UDR   & \checkmark & \checkmark & \checkmark & -          & \checkmark \\ \hline
UPF   & \checkmark & \checkmark & \checkmark & \checkmark & -          \\ \hline
\end{tabularx}
\normalsize
\end{table}

Additionally, Table~\ref{tab:comparativo-5gcs} summarizes the main features available in each of the investigated 5GC platforms. The comparative analysis of the technical capabilities presented in Tables~\ref{tab:mapa-implementacoes-nfv-por-5gcs} and~\ref{tab:comparativo-5gcs} reveals that, although some platforms already offer implementations compliant with 3GPP Release 17, several relevant features remain unavailable across all of them. Given the higher deployment complexity of Magma, free5GC stands out as the most suitable option, offering an excellent balance between ease of deployment and the flexibility required to implement and extend functionalities in multi-RAT environments.

Such information is considered essential for supporting the selection of an appropriate platform for specific scenarios, whether for laboratory experiments, the development of new functionalities, or simulated deployments.

\begin{table}[htb!]
\centering
\scriptsize
\caption{Comparison of open-source 5GC platforms.}
\label{tab:comparativo-5gcs}
\begin{tabular}{|l|c|c|c|c|c|} \hline
& \textbf{Open5GS} & \textbf{free5GC} & \textbf{OAI 5GC} & \textbf{Magma} & \textbf{SD-Core} \\ \hline 
\textbf{Release}    
& 17               
& 15               
& \makecell{16 and 17}         
& -              
& \makecell{16 and 17}\\ \hline
\textbf{Slicing}    
& \checkmark       
& \checkmark       
& \checkmark       
& x              
& \checkmark       \\ \hline
\textbf{Handover}   
& \checkmark       
& \checkmark       
& \checkmark       
& \checkmark     
& \checkmark       \\ \hline
\textbf{Multi-RAT}  
& \makecell{LTE \\ + NR}         
& \makecell{NR \\ + Wi-Fi}       
& x                
& \makecell{LTE + NR \\ + Wi-Fi}       
& \makecell{LTE \\ + NR}         \\ \hline

\end{tabular}
\normalsize
\end{table}

\subsection{Multi-RAT Network Slicing Testbed}

\begin{figure}[H]
\centering
\includegraphics[width=\columnwidth]{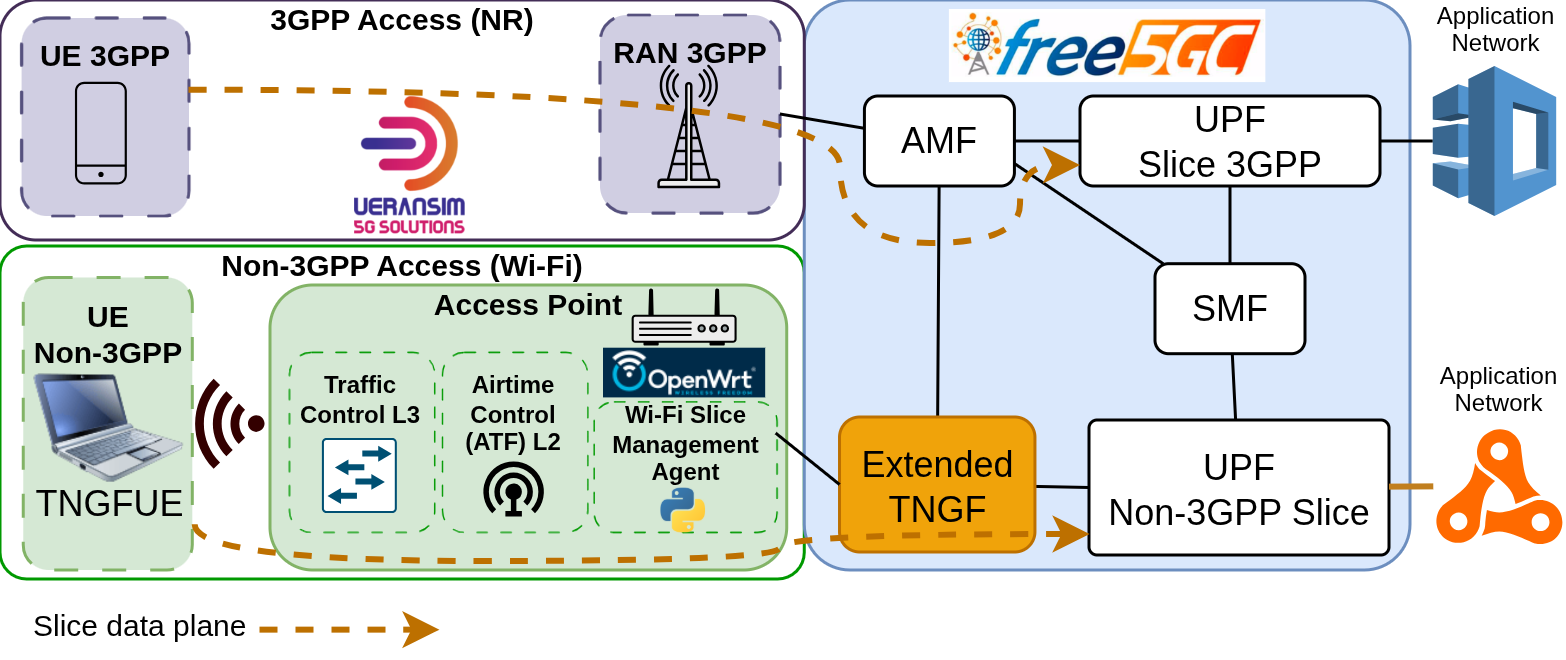}
\caption{High-level view of the multi-RAT network slicing testbed architecture.}
\label{fig:testbed-architecture}
\end{figure}

We depart from the assumption that the required network slicing lifecycle management functions are implemented at either the Operational Support Systems (OSS) or Business Support Systems (BSS) level, at a minimum: Communication Service Management Function (CSMF), the Network Slice Management Function (NSMF), and the Network Slice Subnet Management Function (NSSMF). These functions are essential to initiate the instantiation of network slices in the underlying infrastructure. Fig.~\ref{fig:testbed-architecture} conceptually illustrates the logical structure of the testbed. 

In free5GC, the CSMF translates communication service requirements into technical parameters, which are then managed by the NSMF throughout the NSI lifecycle. The NSSMF, in turn, controls the specific subnets that compose the NSIs, such as RAN, transport, and 5GC. As conceptually illustrated in Fig.~\ref{fig:testbed-architecture}, we extended the legacy \gls{tngf} and NSSMF network functions with new capabilities to afford seamless integration of Wi-Fi networks into NSIs. The enhanced NSSMF becomes responsible for commissioning NSI resources in the Wi-Fi Access Point (AP), while the extended TNGF performs the necessary AP configurations whenever a wireless device attempts to access the system via Wi-Fi. At the Wi-Fi AP, each NSI is mapped to a dedicated transmission queue, identified by a unique Single Network Slice Selection Assistance Information (SST) and Slice Differentiator (SD) tuple. Additional networking parameters are also configured, including:

\begin{itemize}
    \item \textbf{Airtime configuration}: Based on the 5G QoS Identifier (5QI) registered in the 5GC for the NSI associated with the IMSI of a UE (now operating as a Wi-Fi STA), the \gls{atf} mechanism is configured. This mechanism modifies how the Wi-Fi MAC layer distributes transmissions among clients by performing differentiated allocation of TXOPs (Transmission Opportunities) based on QoS Flow Identifier (QFI) values.
 
    \item \textbf{Bandwidth configuration}: Based on the Downlink Aggregate Maximum Bit Rate (DL AMBR) defined for the NSI in the 5GC, a Hierarchical Token Bucket (HTB) queue with limited bandwidth is associated with the UE’s IP address and configured in the corresponding Wi-Fi AP. HTB, part of the Linux kernel queuing subsystem, is widely used in Linux routers and OpenWrt and was employed in this work for traffic control and prioritization.
\end{itemize}

In addition to radio resource management, Wi-Fi operating as a 5G access network can provide secure and reliable connections by inheriting the 5G security framework. This is achieved through a key-derivation hierarchy that links the free5GC to the Non-3GPP access network. During the EAP-5G authentication process, the UE and the AMF generate the $K_{\text{TNGF}}$ key, which is subsequently sent by the AMF to the TNGF. Acting as the security entry point, the TNGF derives and forwards the $K_{\text{TNAP}}$ key to the UE. Both entities use this key to establish encryption and protect over-the-air traffic. Furthermore, a derived $K_{\text{IPsec}}$ key is used to establish a secure IPsec tunnel (NWt interface) between the UE and the TNGF, through which NAS signaling messages are exchanged \cite{3gpp33501_v20_1}.

Once the $K_{\text{IPsec}}$ tunnel is established, the AMF sends the NAS Registration Accept message to the TNGF, which forwards it to the UE, completing the registration procedure. The UE then requests the establishment of a PDU Session over this secure channel. After approval by the AMF and SMF, the TNGF receives the corresponding QoS rules. The TNGF and UE subsequently exchange IKEv2 messages to create a dedicated IPsec Child Security Association (SA) for user-plane traffic. Upon successful configuration and resource reservation, the TNGF sends the PDU Session Establishment Accept message to the UE.

Finally, the TNGF instructs the Wi-Fi AP to associate the UE with the corresponding NSI. From this point forward, user data flows encapsulated in (IP, GRE, PDU) packets, with all communication integrity guaranteed by the 5G security architecture \cite{3gpp33501_v20_1, 3gpp23502_v20_1}.
Some limitations of the proposed solution should be highlighted: (i) the Wi-Fi AP must run on a Linux-based system (e.g., OpenWrt or MikroTik RouterOS) to enable HTB queue manipulation, and its radio driver must support per-connection airtime scheduling; (ii) the UE must support WPA2/WPA3-Enterprise authentication, as it enables 802.1X/EAP-based authentication required for trusted Non-3GPP access integration with the 5GC via TNGF.

\section{Experimental Setup and Test Cases}
\label{sec_setup}
Fig.~\ref{fig:architectural-framework} conceptually illustrates the testbed infrastructure of the experimental testbed, which is centered on the free5GC core network with the proposed extensions to support Wi-Fi as a trusted Non-3GPP access network. The components shown in Fig.~\ref{fig:architectural-framework} are detailed below:

\begin{figure}[htb!]
	\centering
  	\includegraphics[width=\columnwidth]{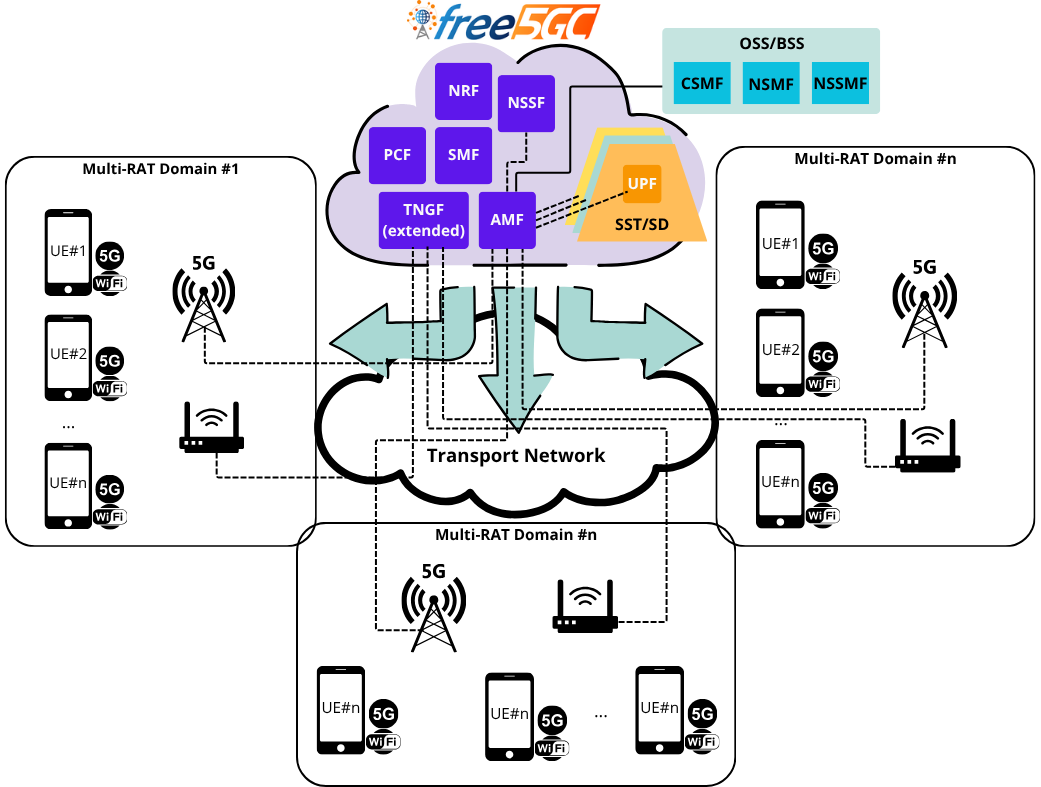}
  	\caption{Architectural framework of the multi-RAT network slicing testbed.}
  	\label{fig:architectural-framework}
\end{figure}

\begin{itemize}
    \item \textbf{3GPP Radio Access Network:} open source state-of-the-art 5G Standalone RAN (gNB) with NGAP and NAS (control plan) features as well as User Plane via GTP protocol.
    \item \textbf{Non-3GPP Radio Access Network:} an Infoway W7030 netbook equipped with an Intel N455 processor, along with Realtek RTL810xE and Qualcomm Atheros AR9285 interfaces, employed to act as a Wi-Fi AP based on OpenWrt version 24.10.1. The AP is Wi-Fi 4 (802.11n, SISO, 20 MHz) with a maximum transmission capacity of 56~Mbps.
    \item \textbf{User Equipments (UEs) or STAs:} Dell Inspiron 3520 and Sony PCG-61A11X notebooks, both running the TNGFUE client to connect to the AP and establish the wireless communication channels.
    \item \textbf{Transport Network:} Typical unmanaged 8-port Gigabit switch.
    \item \textbf{5G Core Network:} free5GC (v4.2.0) is configured on an Ubuntu 22.04 LTS Dell OptiPlex 3080 desktop (Intel i5-10500, 48~GB RAM). Separate UPF instances can be instantiated per NSI, to enforce slice traffic isolation. The core is modified for native-supported TNGF termination, managing registrations and authentications via EAP-5G, and Session Management Function (SMF) session creations for Non-3GPP connections.
\end{itemize}

The source code of the TNGF function was extended to extract, during the UE registration procedure within the 5GC, the DL AMBR information, the 5QI, and the SST/SD parameters associated with the requested NS. Subsequently, an asynchronous message is sent to the AP originating the connection, containing the respective bandwidth limits, ATF priority parameters, and the identification of the NS to which the UE is associated.

Regarding 3GPP access paths, both registration and session establishment follow the standard NAS registration procedures. In the Wi-Fi path, in turn, the procedure consists of three major sequential phases:
\begin{enumerate}
    \item \textbf{IP Connectivity:} The UE connects to the Wi-Fi AP via enterprise WPA2 protocols and receives a local IP address taken from the AP's local DHCP scope, for signaling exchange with TNGF.
    \item \textbf{IKEv2/IPsec SA:} The UE initiates an Internet Key Exchange (IKEv2) signaling procedure with the public-facing IP address of the TNGF. During this exchange, an EAP-5G protocol encapsulation is triggered inside IKEv2 payloads, passing the UE’s 5G NAS registration messages (containing the S-NSSAI) directly to the AMF via the N2 interface.
    \item \textbf{PDU Session Activation:} Once authenticated, the 5GC establishes an IPsec SA child tunnel between the UE and the TNGF. The SMF instructs the UPF to allocate a PDU session IP address from the designated slice network pool, routing all subsequent user plane traffic securely inside a GTP-U tunnel between the TNGF and the UPF.
\end{enumerate}

\subsection{Test Cases}

To validate the multi-RAT network slicing testbed, an experimental scenario was configured using a Wi-Fi NSI with two UEs registered in the 5GC. Each UE was identified by a unique International Mobile Subscriber Identity (IMSI), enabling individualized security policies and differentiated QoS treatment based on 5QI and Downlink Aggregate Maximum Bit Rate (DL AMBR) parameters. The experiment consisted of six operating modes over time. Scenarios 1 and 6 ran under best-effort conditions without network slicing. Scenarios 2 to 5 activated the NSI and applied different combinations of bandwidth limits and traffic treatment parameters. Table~\ref{tab:cenarios_orquestracao} details the configuration of each test case.

\begin{table}[ht]
\centering
\scriptsize
\caption{Experimental scenarios.}
\label{tab:cenarios_orquestracao}

\begin{tabularx}{\columnwidth}{c c c X c c}
\hline
\textbf{Scenario} & \textbf{Interval} & \textbf{Mode} & \textbf{UE1/UE2 Limit} & \makecell{\textbf{ATF} \\ \textbf{UE1/UE2}} & \makecell{\textbf{5QI} \\ \textbf{UE1/UE2}} \\
\hline
1 & 0--50 s    & BE & --            & --         & -- \\
2 & 50--100 s  & NS & 20/20~Mbps    & 512/512    & 3/3 \\
3 & 100--150 s & NS & 40/40~Mbps    & 512/512    & 3/3 \\
4 & 150--200 s & NS & 40/40~Mbps    & 1024/256   & 87/4 \\
5 & 200--250 s & NS & 40/40~Mbps    & 256/1024   & 4/87 \\
6 & 250--300 s & BE & --            & --         & -- \\
\hline
\end{tabularx}
\normalsize
\end{table}

The 5QI values were selected in accordance with Table 5.7.4-1 of the 3GPP TS 23.501 \cite{3gpp_ts_23_501_v19_9_2025}. In Scenarios 2 and 3, both UEs used the same 5QI (5QI 3 for real-time gaming), resulting in homogeneous QoS treatment with differentiation applied only through DL AMBR bandwidth limits. In Scenarios 4 and 5, different 5QIs were assigned (5QI 87 for interactive services and 5QI 4 for non-conversational video) to evaluate traffic prioritization. In Scenario 4, UE1 received higher priority, while in Scenario 5, the priority was assigned to UE2.

\section{Results and Discussion}
\label{sec_results}

Performance evaluation focused on throughput stability, service prioritization, and bandwidth enforcement under concurrent traffic. Traffic was generated using \texttt{iPerf3} with UDP flows from 5GC to UE.

\begin{figure}[h]
\centering
\includegraphics[width=\columnwidth]{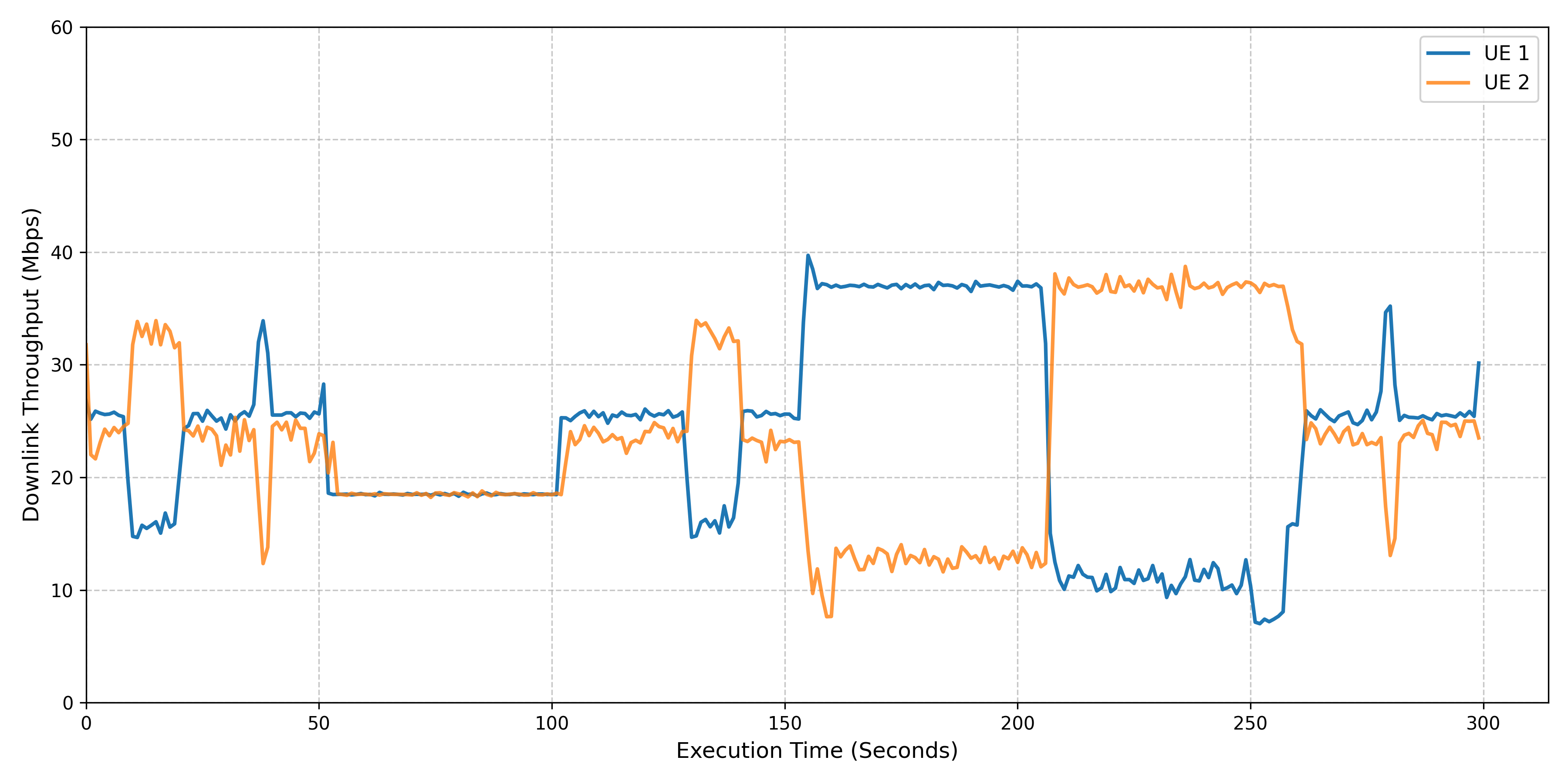}
\caption{Network slicing experiments in Wi-Fi.}
\label{fig:resultados-fatiamento}
\end{figure}

Figure~\ref{fig:resultados-fatiamento} shows the throughput achieved by each UE across the six scenarios. In Scenarios 1 and 6 (best-effort), both UEs exhibited significant throughput variation due to contention for wireless resources. In Scenario~2, the activation of NS with a limit of 20~Mbps for both users and a homogeneous QoS configuration (same 5QI) promotes greater stability in bandwidth distribution and constrains the throughput to the configured value. In Scenario~3, the limit is increased to 40~Mbps, allowing the UEs to request rates exceeding the AP capacity. As expected, this condition generates contention for transmission resources, similarly to the best-effort scenario.

In Scenarios 4 and 5, the combination of Airtime Fairness (ATF) weights and different 5QI values enabled clear traffic differentiation. In Scenario~4, UE1 receives a higher MAC Layer priority, reflected in a higher average throughput, while UE2 operates with a smaller share of the available resources. In Scenario~5, this prioritization is reversed, causing UE2 to receive preferential treatment.

The results demonstrate the capability of the proposed integrated QoS mechanism to efficiently apply dynamic resource management and QoS differentiation policies, even in a shared Wi-Fi environment. Overall, the findings validate the effectiveness of dynamic resource management in Wi-Fi networks integrated with the 5GC.

\section{Conclusions}
\label{sec_conclusoes}

This work presents a prototype investigation for the seamless integration of 5G and WLAN networks, adhering to standards defined by the 3GPP and the Wi-Fi Alliance. The research demonstrated that it is possible to extend the NS paradigm to Non-3GPP domains, enabling devices without native 5G support to benefit from its security, management, and isolation capabilities.

The experimental results, obtained through a testbed integrating free5GC and OpenWrt, demonstrated that the extension of the TNGF enabled effective control of Wi-Fi network transmission resources. The synthesis of the findings reveals that the use of QoS parameters from the 5GC (such as 5QI and DL AMBR) enabled the dynamic management of bandwidth (through HTB queues) and transmission time (through Airtime Fairness), ensuring traffic stability and prioritization in a manner analogous to native 5G mechanisms.

Despite the achieved advancements, the study identifies as a limitation the need for an even more dynamic and comprehensive integration capable of meeting the more complex requirements of large-scale heterogeneous networks. As future work, the investigation of automated orchestration mechanisms capable of responding in real time to extreme load variations is recommended, as well as the exploration of machine learning techniques to optimize resource allocation.

\bibliographystyle{IEEEtran}
\bibliography{bibliografia}

\end{document}